\documentclass[twocolumn, tighten, times]{aastex631}
\shorttitle{Hydrodynamic Response of Mildly Evolved Common Envelope Donors}
\shortauthors{Hutchinson-Smith et al.}
\usepackage{xcolor}

\begin{document}

\title{Hydrodynamic Response of Mildly Evolved Common Envelope Donors in Luminous Red Novae}

\correspondingauthor{Tenley Hutchinson-Smith}
\email{tenley@ucsc.edu}

\author[0000-0002-3472-2453]{Tenley Hutchinson-Smith}
\affiliation{Department of Astronomy and Astrophysics,
University of California,
Santa Cruz, CA,  95064, USA}

\author[0009-0006-4675-7596]{Angela A. Twum}
\affiliation{Department of Astronomy and Astrophysics,
University of California,
Santa Cruz, CA,  95064, USA}

\author[0000-0003-2558-3102]{Enrico Ramirez-Ruiz}
\affiliation{Department of Astronomy and Astrophysics,
University of California,
Santa Cruz, CA,  95064, USA}

\begin{abstract}
Luminous red novae trace unstable binary interactions in which common-envelope evolution can produce either a stellar merger or a surviving binary following envelope ejection. Recent population studies suggest that a substantial fraction of these systems originate from mildly evolved donors whose structures occupy an intermediate regime between simplified polytropic envelopes and the highly stratified giant stars commonly studied in classical common-envelope calculations. We present a suite of three-dimensional hydrodynamic simulations of mildly evolved donors interacting with embedded companions spanning a range of mass ratios and central density concentrations. We show that the hydrodynamic evolution is strongly regulated by the donor central concentration, parameterized by the ratio $\rho_c/\bar{\rho}$. Donors with similar values of $\rho_c/\bar{\rho}$ exhibit qualitatively similar inspiral morphologies and mass-ejection histories despite substantial differences in stellar mass and radius. Systems with relatively modest central concentration evolve through rapid inspiral dominated by local orbital-energy deposition, while more centrally concentrated donors develop prolonged expansion-driven phases in which shocks, circulation flows, and large-scale envelope motions efficiently redistribute deposited energy and angular momentum throughout the star. In this regime, the envelope itself becomes dynamically important in driving continued mass loss long after the rapid plunge-in phase has slowed substantially. These results challenge semi-analytic models of luminous red novae that assume nearly instantaneous envelope ejection and estimate ejection efficiencies using simple orbital-energy balance arguments. Instead, our simulations suggest that the observable diversity of luminous red novae may depend not only on the total ejecta mass, but also on the temporal structure of the outflow, including the transition between impulsive envelope ejection and prolonged wind-like or multi-stage mass-loss evolution.
\end{abstract}

\keywords{binaries: close --- stars: evolution --- stars: interiors --- hydrodynamical simulations: common envelope evolution --- transient sources: luminous red novae}
    
\section{Introduction} 
A common-envelope interaction occurs when one star in a binary expands and engulfs its companion, initiating rapid orbital inspiral and strong hydrodynamic coupling between the embedded companion and the donor envelope \citep{pazcynski1976,apjac6269bib101,apjac6269bib110,apjac6269bib30,2013A&ARv..21...59I, apjac6269bib93}. 
These interactions can produce two qualitatively different outcomes: successful envelope ejection leading to binary survival, or continued inspiral culminating in stellar merger \citep{apjac6269bib101,Podsiadlowsk2001,apjac6269bib78,2018ApJ...863....5M,apjac6269bib38}.
Common-envelope interactions are widely believed to power luminous red novae (LRNe), a class of optical and infrared transients associated with eruptive mass ejection during unstable binary interaction and coalescence \citep{apjac6269bib55,apjac6269bib65,apjac6269bib9,soker2003,apjac6269bib45,apjac6269bib114,apjac6269bib46,apjac6269bib100,apjac6269bib7,apjac6269bib10,apjac6269bib75,apjac6269bib107,apjac6269bib6,2026arXiv260517005K}.

Recent binary-population synthesis studies have shown that a substantial fraction of LRN-producing systems originate from mildly evolved donors undergoing unstable mass transfer during the early stages of post-main-sequence expansion \citep{2026arXiv260210211T}. 
In this regime, the inspiral is expected to proceed toward merger rather than successful envelope ejection and long-term binary survival \citep[e.g.,][]{2024ApJ...977..196H}. 
These donors occupy an intermediate structural regime between zero-age main-sequence stars and stars with highly stratified envelopes, making them particularly valuable for studying the hydrodynamic response of stars to embedded companions. 
Their envelopes are sufficiently extended to undergo strong dynamical interaction while remaining only moderately centrally concentrated, placing these systems in a transition regime where the global hydrodynamic response remains strongly coupled to the stellar interior.

This regime is also particularly well suited for three-dimensional hydrodynamical calculations \citep[e.g., ][]{2020ApJ...901...44W,2024ApJ...977..196H,2024ApJ...971..132E}. Stars with highly stratified envelopes develop extreme density contrasts that are computationally challenging to resolve and tend to localize the hydrodynamic response to the outermost layers of the star \citep[e.g.,][]{2013A&ARv..21...59I,2017ApJ...835..282M,2023LRCA....9....2R}. Mildly evolved donors instead exhibit more moderate values of the ratio of central to average density, $\rho_c/\bar{\rho}$, making them both more computationally tractable and more conducive to efficiently coupling the perturbation generated by the embedded companion to the stellar interior \citep[e.g.,][]{2025ApJ...979L..11E}. As a result, relatively small changes in stellar structure can produce large differences in inspiral dynamics, shock formation, and mass ejection.

This sensitivity to stellar central concentration has also emerged in studies of tidal disruption events \citep[e.g.,][]{2019ApJ...882L..25L,2020ApJ...905..141L,2025ApJ...990L...7B,2025ApJ...979...40L}. Three-dimensional simulations have shown that many disruption outcomes can be organized primarily by the ratio of central to average density, $\rho_c/\bar{\rho}$, enabling the construction of numerical libraries parameterized by stellar central concentration rather than stellar mass and radius alone \citep{2020ApJ...905..141L}. In contrast to tidal disruption events, where the perturbation is externally driven, common-envelope interactions involve a lower-mass companion inspiraling within the donor envelope, depositing energy and angular momentum throughout the stellar interior \citep{2023LRCA....9....2R}. Here we investigate how the hydrodynamic evolution of common-envelope inspiral varies across donors spanning a range of $\rho_c/\bar{\rho}$, with particular emphasis on whether systems with similar central concentration exhibit comparable inspiral morphologies and mass-ejection histories despite differences in stellar mass and radius \citep{2020ApJ...905..141L,2020ApJ...899...77E}. We focus on mildly evolved donors undergoing merger-driven inspiral with embedded companions spanning a range of mass ratios motivated by the population synthesis models of \citet{2026arXiv260210211T}.

The efficiency and duration of envelope ejection during common-envelope evolution remain highly uncertain \citep{1988ApJ...329..764L,1998ApJ...500..909S,2003RMxAC..15...34D,2012ApJ...744...52P,apjac6269bib93}. In particular, it remains unclear how efficiently orbital energy couples to the envelope and over what timescales mass loss proceeds. Additional processes, including recombination-powered outflows \citep{2013A&ARv..21...59I} and jet feedback from the embedded companion \citep{2015ApJ...800..114S,2023ApJ...954..143D,2024Galax..12...33S}, may further accelerate envelope removal or alter the late-time evolution of the interaction.

Our goal is to determine how donor structure regulates inspiral dynamics and mass ejection during the hydrodynamic evolution leading to luminous red novae. Many models used to interpret the light curves of luminous red novae assume that the bulk of the envelope mass is ejected impulsively over only a few dynamical timescales following the onset of inspiral \citep{2013Sci...339..433I,2017ApJ...835..282M}. Whether this assumption applies across the full diversity of LRN progenitors remains uncertain, particularly for donors whose internal structure may regulate the timescale and character of mass ejection in ways not captured by simple orbital-energy balance arguments.  

To explore this connection, we utilize the grid of donor stars from \citet{2026arXiv260210211T}, constructed using the one-dimensional stellar evolution code \texttt{MESA} \citep{apjac6269bib79,apjac6269bib80,apjac6269bib81,paxton2018,apjac6269bib82,2023ApJS..265...15J}, to survey the landscape of central concentration among mildly evolved donors. From this grid, we select donor models spanning a representative range of $\rho_c/\bar{\rho}$ and simulate their common-envelope evolution using \texttt{FLASH} \citep{2000ApJS..131..273F}, a three-dimensional adaptive mesh refinement hydrodynamics code. We additionally perform comparison calculations involving donors with similar values of $\rho_c/\bar{\rho}$ but substantially different masses and evolutionary states to test the extent to which central concentration governs the hydrodynamic evolution. By following the inspiral self-consistently in three dimensions, we investigate how donor central concentration regulates the inspiral dynamics, hydrodynamic response, and temporal distribution of mass ejection during common-envelope evolution.

The paper is structured as follows. Section 2 introduces the stellar evolution and population synthesis models of \citet{2026arXiv260210211T} used to map the landscape of donor central concentration, with particular emphasis on the regime occupied by mildly evolved progenitors. Section 3 presents the hydrodynamical simulations, introducing the primary donor models and the \texttt{FLASH} setup, and explores the inspiral dynamics and mass-ejection histories across a range of companion mass ratios. Section 4 discusses the implications of our results for common-envelope evolution and luminous red novae. Appendix~\ref{comp} presents additional comparison simulations involving donors with similar values of $\rho_c/\bar{\rho}$ but different masses and evolutionary states, while Appendix~\ref{cesim} summarizes the donor masses and radii adopted in previous global common-envelope simulations and places the present models within the broader landscape of common-envelope calculations.

\begin{figure*}
    \centering
    \includegraphics[width=1\linewidth]{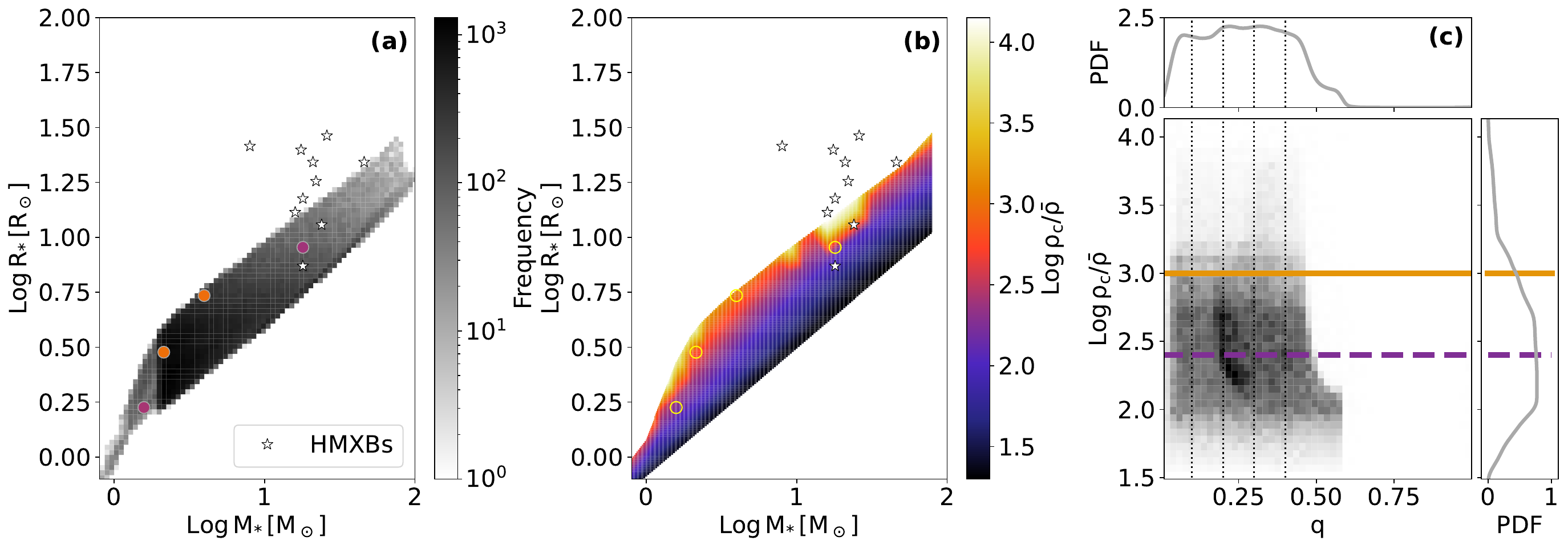}
    \caption{Binary interactions involving unstable mass transfer obtained from the binary population synthesis calculations of \citet{2026arXiv260210211T}. Panel {\it a}: Distribution of mildly evolved (Zone I) donors in the mass-radius plane. Panel {\it b}: Distribution of the ratio of central to average density, $\rho_c/\bar{\rho}$, across the mass-radius plane for the mildly evolved donor population. In panels {\it a} and {\it c}, symbols indicate the subset of donor models selected for the three-dimensional hydrodynamical simulations. Panel {\it c}: Joint distribution of binary mass ratio, $q$, and donor central concentration, $\rho_c/\bar{\rho}$, for mildly evolved common-envelope donors. The masses and radii of eclipsing high-mass x-ray binaries (HMXBs) studied by \citet{2015A&A...577A.130F} are shown as star symbols. The simulated models (orange and purple) span the characteristic range of structural and binary properties occupied by the majority of mildly evolved common-envelope donors.}
    \label{fig:1}
\end{figure*}

\section{Demographics of Mildly Evolved Common-Envelope Donors}

We adopt the binary population synthesis framework presented by \citet{2026arXiv260210211T}, which combines rapid binary evolution calculations using \texttt{COMPAS} \citep{Setvenson2017,TeamCompas2022,TeamCompas2025} with detailed stellar structure models to follow interacting binaries undergoing unstable mass transfer and common-envelope evolution. In this work, we specifically adopt the binary population model based on the orbital distributions inferred for massive binaries by \citet{Sana2012}, referred to as the {\tt SANA} model in \citet{2026arXiv260210211T}. This model includes distributions in primary mass, orbital period, and mass ratio and produces the largest fraction of interacting systems originating from mildly evolved donors among the population models explored in that work. Systems are evolved until the onset of unstable mass transfer, at which point they are classified according to the evolutionary state and internal structure of the donor star. Of particular relevance here are the mildly evolved donors identified as Zone I systems in \citet{2026arXiv260210211T}. These correspond to binaries that initiate unstable mass transfer shortly after leaving the main sequence, before developing highly stratified envelopes. Under the \citet{Sana2012} binary parameter distributions, mildly evolved donors constitute approximately 27.4\% of the predicted common-envelope interactions associated with luminous red novae, making them one of the dominant progenitor channels. Furthermore, the population synthesis models indicate that the overwhelming majority of these systems undergo only a single common-envelope interaction \citep{2026arXiv260210211T}, implying that the donor structures considered here are unlikely to have been substantially altered by previous mass-transfer episodes. This lends further support to the use of single-star stellar evolution models in constructing the donor structures adopted in this work.

Mildly evolved donors occupy a relatively narrow evolutionary window in which the stellar envelope has expanded beyond the zero-age main sequence but has not yet developed the extreme density stratification characteristic of later evolutionary stages. Their envelopes remain only moderately centrally concentrated, allowing the hydrodynamic response to remain globally coupled to the stellar interior. This makes these systems particularly valuable for studying the response of common-envelope donors to embedded inspiral. 

To characterize the internal structure of these donors, we compute the ratio of central to average density, $\rho_ c/\bar{\rho}$, across the donor population. Figure~\ref{fig:1} shows the distribution of mildly evolved donors in the mass-radius plane together with the corresponding distribution of $\rho_c/\bar{\rho}$. Mildly evolved donors populate a transition regime where relatively small changes in stellar evolution produce substantial changes in stellar central concentration. Most of these donors occupy an intermediate structural regime with $10^2 \lesssim \rho_c/\bar{\rho} \lesssim 10^3$. For comparison, a $\Gamma=5/3$ polytrope has $\rho_c/\bar{\rho}=5.8$, while a $\Gamma=4/3$ polytrope has $\rho_c/\bar{\rho}=73$ \citep{2020ApJ...905..141L}, whereas highly stratified giant envelopes can reach $\rho_c/\bar{\rho}\gtrsim 10^6$ \citep{2012ApJ...757..134M}

The donor structures are calculated using the one-dimensional stellar evolution code MESA \citep{apjac6269bib79,apjac6269bib80,apjac6269bib81,paxton2018,apjac6269bib82,2023ApJS..265...15J}  following the methodology described in {\cite{2026arXiv260210211T}}. To capture the diversity of common-envelope donors associated with luminous red novae, \cite{2026arXiv260210211T} constructed extensive grids of stellar evolution models spanning initial masses from $0.08$ to $100 M_\odot$ at solar metallicity ($Z=0.02$). Low- and intermediate-mass stars ($0.08$-$7.9 M_\odot$) were evolved using the MESA \texttt{7M\_prems\_to\_AGB} test suite, which follows stellar evolution from the pre-main-sequence phase through the asymptotic giant branch. These calculations adopt a mixing length parameter of $1.73$ and include Reimers and Blocker prescriptions for RGB and AGB mass loss, respectively. Intermediate- and high-mass stars were evolved using the \texttt{12M\_pre\_ms\_to\_core\_collapse} configurations, following stellar evolution from the zero-age main sequence to core collapse.

From this library, we select mildly evolved systems undergoing unstable mass transfer shortly after leaving the main sequence. The corresponding donor structures are extracted from the MESA stellar evolution calculations at evolutionary stages representative of the onset of unstable interaction and are subsequently mapped into the three-dimensional hydrodynamical simulations described in Section~\ref{flash}. For each donor, we extract the radial density, pressure, temperature, and composition profiles needed to initialize the hydrodynamical calculations.

The subset of donor models selected for the three-dimensional simulations is indicated by open symbols in Figure~\ref{fig:1}. These models span the characteristic range of $\rho_c/\bar{\rho}$ occupied by the majority of mildly evolved progenitors while sampling a broad range of donor masses and companion mass ratios. The simulation suite is designed to isolate the role of stellar central concentration by comparing systems with substantially different donor masses but similar values of $\rho_c/\bar{\rho}$, as well as systems with fixed donor structure and varying companion mass ratio. This allows us to systematically explore how stellar central concentration regulates inspiral dynamics, shock formation, and mass ejection during common-envelope evolution. Details of the three-dimensional hydrodynamical calculations are presented in Section~\ref{flash}. For comparison, Figure~\ref{fig:1} also includes observed high-mass X-ray binaries hosting neutron stars, which are expected to evolve into common-envelope systems once the massive donor fills its Roche lobe and unstable mass transfer leads to the engulfment of the neutron star within the expanding envelope \citep{2024ApJ...977..196H}.

\section{Hydrodynamical Simulations}
\label{flash}

\subsection{{\tt FLASH} Setup}
The one-dimensional donor structures generated with MESA are imported into the three-dimensional hydrodynamical framework using {\tt FLASH} version 4.32 \citep{2000ApJS..131..273F}. For each donor, the radial density, pressure, temperature, and composition profiles are interpolated onto a Cartesian adaptive mesh refinement grid to initialize the stellar envelope in hydrostatic equilibrium. The hydrodynamical evolution is computed using the numerical framework developed for stellar disruption and merger calculations by \citet{2009ApJ...705..844G} and \citet{2013ApJ...767...25G}, later extended to follow detailed thermodynamic and compositional evolution in three dimensions \citep{2017ApJ...841..132L,2020ApJ...905..141L,2020ApJ...901...44W,2025ApJ...990L...7B}. The simulations employ the extended Helmholtz equation of state \citep{2000ApJS..126..501T}, which allows us to consistently evolve the thermodynamic state of the gas throughout the inspiral. 

Our simulation suite is designed to isolate the role of donor central concentration in regulating common-envelope hydrodynamics. The primary calculations focus on two representative donor families spanning contrasting values of $\rho_c/\bar{\rho}$ (Figure~\ref{fig:1}). The first consists of an $18M_\odot$ donor with moderate central concentration, $\rho_c/\bar{\rho}=251$, while the second consists of a $2.14M_\odot$ donor with substantially higher central concentration, $\rho_c/\bar{\rho}=962$. For each donor, we consider embedded companions with mass ratios $q \equiv M_\bullet/M_\ast$ between $0.1$ and $0.4$, where $M_\ast$ is the mass of the donor and $M_\bullet$ is the mass of the inspiraling companion. This allows us to explore how inspiral dynamics and mass ejection vary with binary mass ratio at fixed donor structure. Additional comparison calculations involving donors with similar values of $\rho_c/\bar{\rho}$ but different masses and evolutionary states are presented in Appendix~\ref{comp}.

\begin{figure*}
\centering
\includegraphics[scale = 0.55]{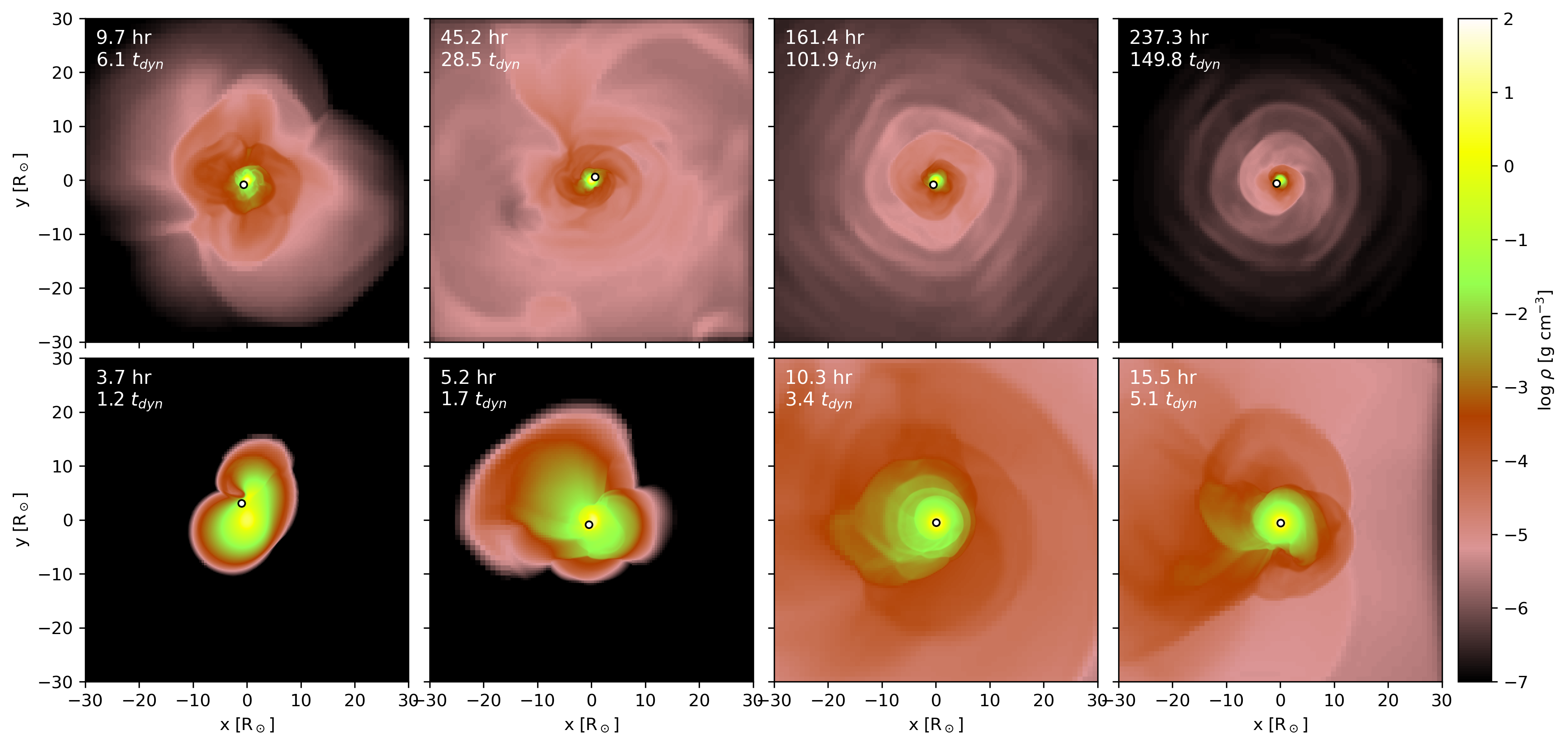}
\caption{
Time evolution of the mid-plane density  for the two representative $q=0.1$ simulations. The more centrally concentrated $2.14M_\odot$ donor ({\it top} panels) with $\rho_c/\bar{\rho}=962$ evolves over tens of dynamical timescales and settles into a quasi-stationary orbit while the donor envelope undergoes substantial expansion. Mass loss continues throughout this extended phase for hundreds of dynamical timescales, ultimately leading to the ejection of most of the donor envelope. The moderately concentrated $18M_\odot$ donor ({\it bottom} panels) with $\rho_c/\bar{\rho}=251$ undergoes rapid inspiral over only a few dynamical timescales, after which the orbital separation decreases more gradually as the embedded companion settles near the core radius of the donor.
}
\label{fig:hydro}
\end{figure*}

Before introducing the embedded companion, each donor model is first mapped into {\tt FLASH} and evolved in isolation for several donor dynamical timescales. The simulations are performed in a three-dimensional Cartesian domain with an $8^3$ block grid structure and adaptive mesh refinement. The finest refinement level corresponds to a minimum cell size of $\lesssim R_\ast/10^2$, where $R_\ast$ is the radius of the donor, consistent with the resolution required to achieve convergence in both the inspiral trajectory and mass-ejection evolution \citep{2020ApJ...901...44W,2020ApJ...905..141L, 2024ApJ...977..196H}. During the relaxation procedure, the embedded companion is introduced as a point mass initially placed at rest near the donor surface. Its orbital velocity is then gradually increased until reaching the local circular Keplerian value, allowing the donor envelope to adjust smoothly to the presence of the companion. Once the relaxation phase is complete, the donor has settled into hydrostatic equilibrium on the computational grid and the subsequent inspiral trajectory of the embedded companion is evolved self-consistently \citep{2020ApJ...901...44W}. Additional details regarding the numerical implementation, relaxation procedure, and hydrodynamic scheme can be found in \citet{2024ApJ...977..196H} and \citet{2024ApJ...971..132E}.

To further test the extent to which the hydrodynamic evolution is governed by stellar central concentration rather than donor mass alone, we perform additional comparison calculations involving donors with similar values of $\rho_c/\bar{\rho}$ but different masses and evolutionary states. These include a $1.58M_\odot$ donor with $\rho_c/\bar{\rho}=260$, comparable to the $18M_\odot$ donor, and a $3.97M_\odot$ donor with $\rho_c/\bar{\rho}=940$, comparable to the $2.14M_\odot$ donor. For these comparison models, we focus on the $q=0.1$ cases, which are presented in Appendix~\ref{comp}. These calculations allow us to directly assess whether systems with similar central concentration exhibit similar hydrodynamic evolution despite substantial differences in donor mass.

\begin{figure*}
\centering
\includegraphics[scale = 0.6]{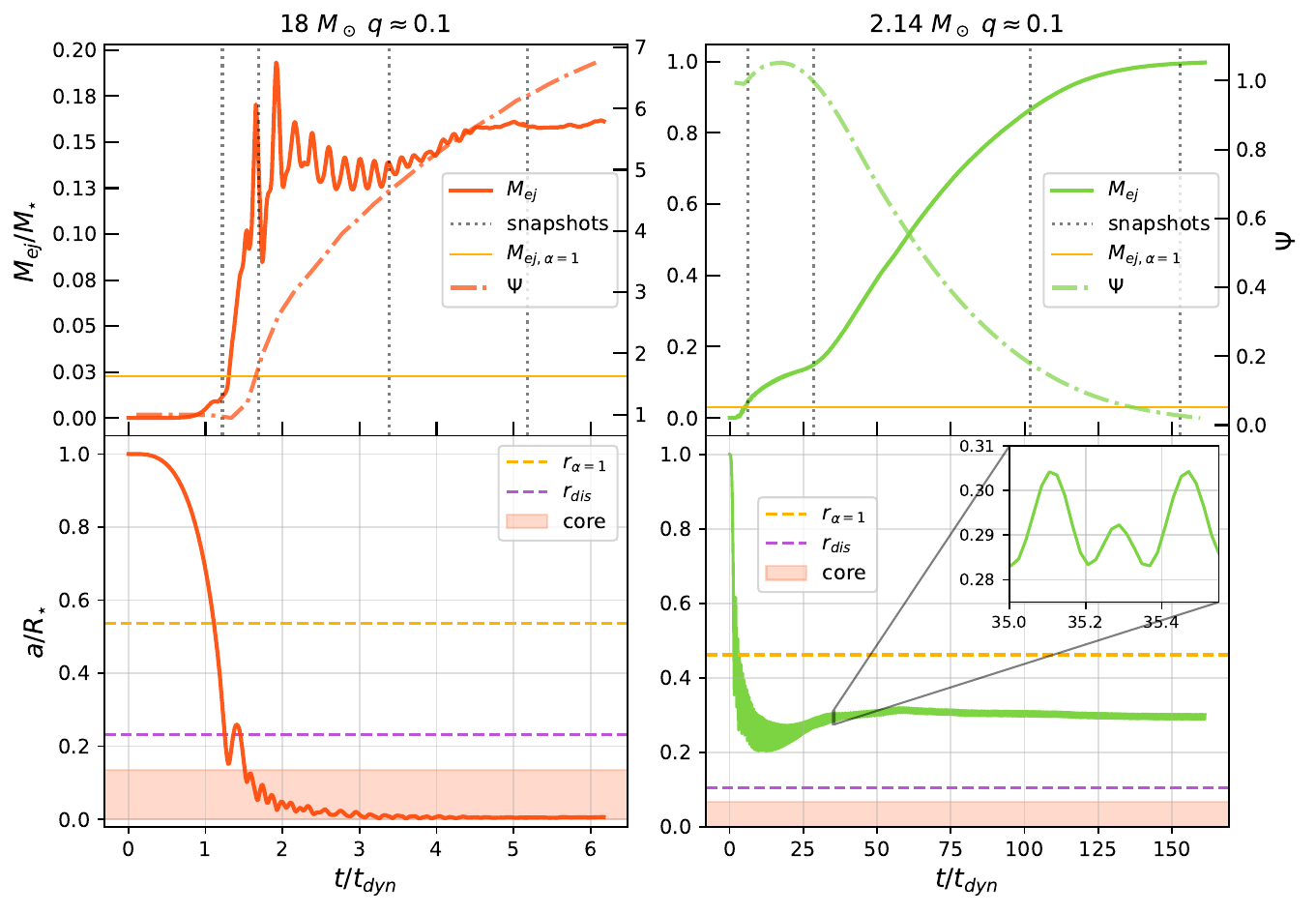}
\caption{Time evolution, in units of the dynamical timescale of the  donor, representative $q=0.1$ simulations for the moderately concentrated $18M_\odot$ donor with $\rho_c/\bar{\rho}=251$ ({\it left}) and the high-central-concentration $2.14M_\odot$ donor with $\rho_c/\bar{\rho}=962$ ({\it right}). The top panels show the cumulative ejected mass, normalized to the initial donor mass, together with the evolution of the enclosed-mass response parameter $\Psi$. Horizontal solid lines indicate the ejecta mass predicted by the standard $\alpha$-formalism assuming $\alpha=1$. Vertical dotted lines mark the times corresponding to the hydrodynamical snapshots shown in Figure~\ref{fig:hydro}. The bottom panels show the orbital separation of the embedded companion normalized to the donor radius, $a/R_\ast$. The horizontal yellow dashed line indicates the characteristic radius at which the released orbital energy equals the binding energy of the overlying envelope in the $\alpha=1$ formalism, while the purple dashed line marks the tidal disruption radius assuming a zero-age main-sequence companion \citep{2026arXiv260210211T}. The pink shaded region denotes the donor's core region.}
\label{fig:dem}
\end{figure*}

\subsection{Dynamical Inspiral at Fixed Companion Mass Ratio}
We first compare the hydrodynamic evolution of the two representative donor families at fixed companion mass ratio, $q=0.1$, thereby isolating the role of donor structure while holding the relative companion mass fixed. Figure~\ref{fig:hydro} compares the mid-plane density evolution of both systems at comparable times measured in units of the donor dynamical timescale. This comparison highlights how increasing donor central concentration alters the global hydrodynamic response during inspiral, leading to qualitatively different shock structures, envelope expansion, and mass-ejection behavior.

The two calculations exhibit qualitatively different inspiral and mass-ejection histories. In the moderately concentrated $18M_\odot$ donor ({\it bottom} panels In Figure~\ref{fig:hydro}), the embedded companion inspirals rapidly through the envelope over only a few donor dynamical timescales. Following this initial plunge, the orbital separation continues to decrease gradually as the companion settles near the core radius of the donor. As the companion inspirals toward the core, it deposits orbital energy and angular momentum into the donor envelope through shocks and spiral density waves, unbinding only a modest fraction of the envelope material. Most of the hydrodynamic response therefore occurs during the initial inspiral phase, after which the large-scale donor structure evolves comparatively little.

By contrast, the more centrally concentrated $2.14M_\odot$ donor ({\it top} panels in Figure~\ref{fig:hydro}) undergoes a much more prolonged evolution. The companion inspirals over tens of donor dynamical timescales before settling into a quasi-stationary orbit. Soon after the initial inspiral phase is completed, the donor envelope begins to expand on progressively larger scales, driving sustained mass loss for hundreds of dynamical timescales and ultimately ejecting most of the envelope. In this case, the slowing of the inspiral marks not the end of the hydrodynamic response, but the onset of a prolonged expansion-driven mass-loss phase.

We quantify these  evolutionary differences in Figure~\ref{fig:dem}, which shows the unbound mass and companion separation as functions of time, in units of the donor dynamical timescale,  for both $q=0.1$ simulations. We follow \citet{2024ApJ...977..196H} and define the ejected mass as gas with positive total specific energy, computed as the sum of the specific kinetic and gravitational potential energies, excluding the internal energy contribution.  To quantify the hydrodynamic response of the donor envelope, we introduce the dimensionless enclosed-mass response parameter,
\begin{equation}
\Psi \equiv \frac{M(<a)}{M_0(<a)},
\end{equation}
where $M(<a)$ is the donor mass enclosed within the instantaneous orbital separation, $a$, of the embedded companion and $M_0(<a)$ is the enclosed mass at the same radius in the original unperturbed donor profile. This quantity directly measures how the donor redistributes mass during the inspiral. Values of $\Psi<1$ correspond to envelope expansion and rarefaction interior to the companion orbit, while $\Psi>1$ indicates compression or contraction.

\begin{figure*}
\centering
\includegraphics[scale = 0.34]{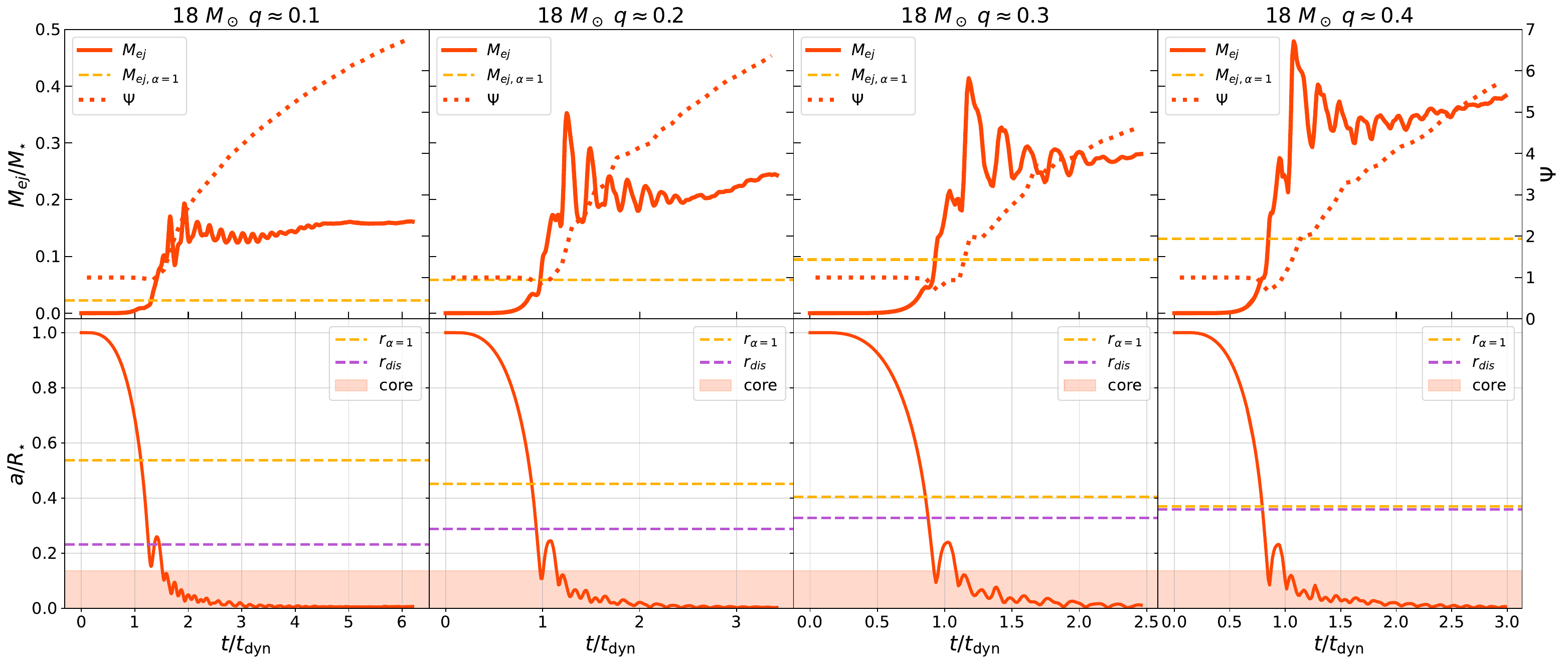}
\caption{Time evolution of the moderately concentrated $18M_\odot$ donor with $\rho_c/\bar{\rho}=251$ for companion mass ratios spanning $q=0.1$--$0.4$. The panels show the cumulative ejected mass, enclosed-mass response parameter $\Psi$, and orbital separation as functions of time in units of the donor dynamical timescale. All models exhibit qualitatively similar inspiral evolution, with the embedded companion rapidly inspiraling toward the donor core over only a few dynamical timescales. The total ejected mass increases systematically with companion mass ratio, although the ratio between the simulated ejecta mass and the $\alpha=1$ prediction does not increase with $q$. Line styles and additional quantities are defined as in Figure~\ref{fig:dem}.}
\label{fig:18q}
\end{figure*}

The evolution of $\Psi$ provides a useful diagnostic for understanding the global hydrodynamic response of the donor and helps explain the markedly different mass-ejection histories of the two systems. The $18M_\odot$ donor ({\it left} panels in Figure~\ref{fig:dem}) ejects material primarily during the initial rapid inspiral and subsequently reaches an approximately steady ejecta mass. Throughout this evolution, $\Psi$ remains greater than unity, indicating that the donor contracts as the companion inspirals and continues to compress until the embedded object settles near the donor core. In this regime, the interaction primarily perturbs the local envelope surrounding the companion, depositing orbital energy and angular momentum into a relatively confined region and unbinding only a modest fraction of the donor mass.

By contrast, the $2.14M_\odot$ donor ({\it right} panels in Figure~\ref{fig:dem}) exhibits an initial contraction phase during the early dynamical inspiral, followed by a sustained decrease in $\Psi$ once the inspiral slows. This transition marks the onset of large-scale envelope expansion, during which the donor continues to unbind material over hundreds of dynamical timescales. As the envelope expands and redistributes mass on progressively larger scales, the ejecta mass continues to increase long after the initial inspiral phase has ended, ultimately leading to the ejection of most of the donor envelope. 

Also shown in Figure~\ref{fig:dem} is the radius at which the embedded companion is expected to undergo tidal disruption. Following \citet{2026arXiv260210211T}, we assume disruption occurs once the inspiraling companion encounters an enclosed donor density comparable to roughly half of its own average density, as expected for a zero-age main-sequence accretor. Once this condition is reached, we assume the companion is tidally disrupted, halting further inspiral and terminating additional orbital energy deposition into the envelope.

For comparison, we also indicate the characteristic radius predicted by the standard $\alpha$-formalism \citep{Webbink1984}, assuming $\alpha=1$, at which the released orbital energy equals the binding energy of the overlying envelope. In both models, this characteristic radius is reached promptly during the initial dynamical inspiral phase, broadly consistent with expectations from simple orbital-energy balance arguments applied to a frozen stellar structure. However, the subsequent hydrodynamic evolution alters the mass-ejection history substantially. While the moderately concentrated donor ejects only a modest fraction of its envelope after reaching this stage, the highly concentrated donor undergoes continued large-scale expansion that drives prolonged mass loss and ultimately unbinds most of the envelope. These calculations demonstrate that the hydrodynamic response of the donor can significantly modify the mass ejection relative to predictions based solely on instantaneous orbital-energy deposition.

\subsection{Dependence on Companion Mass Ratio}
We now investigate how the common-envelope evolution changes with companion mass ratio for fixed donor structure. Figure~\ref{fig:18q} shows the evolution of the moderately concentrated $18M_\odot$ donor with $\rho_c/\bar{\rho}=251$ for companion mass ratios spanning $q=0.1$--$0.4$.

All simulations exhibit qualitatively similar inspiral evolution. In each case, the embedded companion rapidly inspirals through the donor envelope over only a few dynamical timescales before settling near the donor core. The inspiral trajectories remain remarkably similar despite the factor of four variation in companion mass, indicating that the global hydrodynamic evolution is largely set by the donor structure. In all models, the inspiral proceeds well beyond the characteristic $\alpha=1$ radius predicted by the standard energy formalism.

\begin{figure*}
\centering
\includegraphics[scale = 0.34]{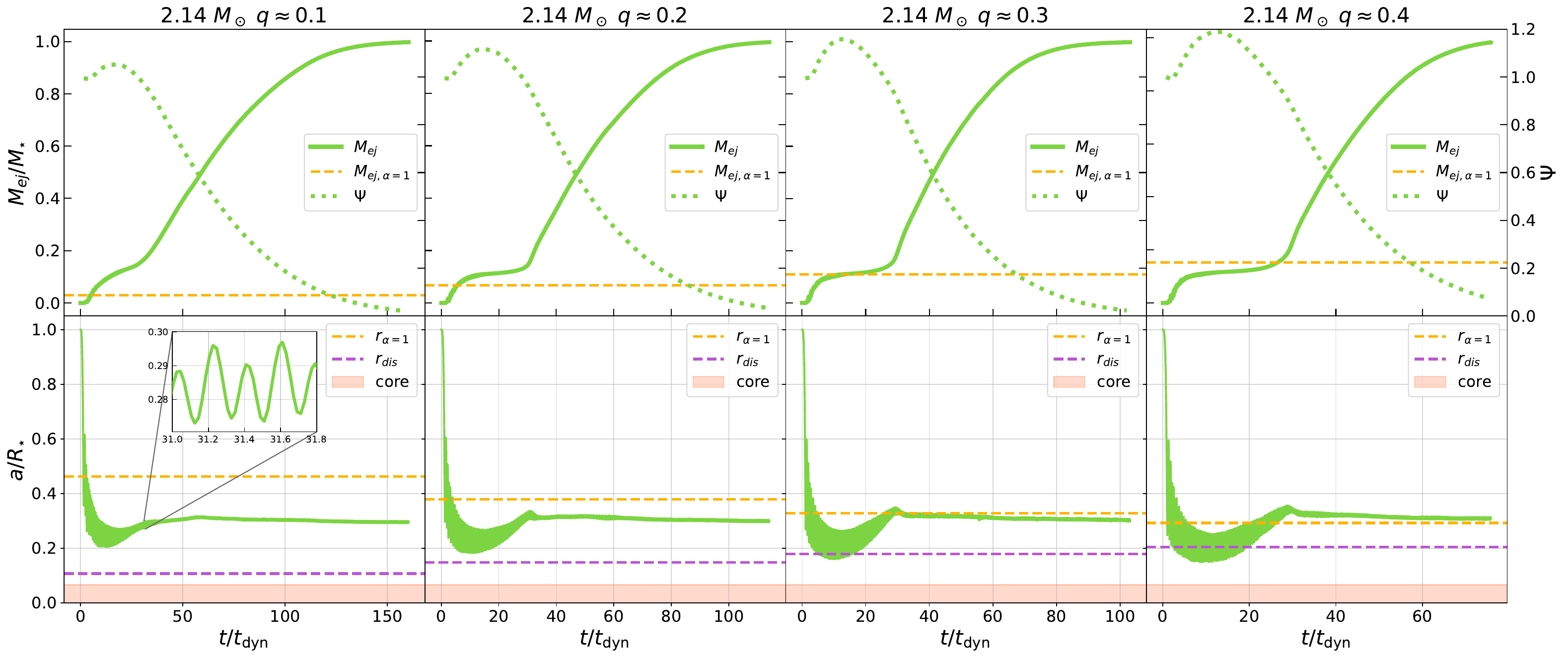}
\caption{Time evolution of the high-central-concentration $2.14M_\odot$ donor with $\rho_c/\bar{\rho}=962$ for companion mass ratios spanning $q=0.1$--$0.4$. The panels show the cumulative ejected mass, enclosed-mass response parameter $\Psi$, and orbital separation as functions of time in units of the donor dynamical timescale. In all models, the inspiral undergoes an initial dynamical contraction phase followed by a transition to prolonged envelope expansion once the companion settles into a quasi-stationary orbit. The subsequent evolution is dominated by sustained hydrodynamic expansion and continued mass loss over tens to hundreds of dynamical timescales, ultimately leading to the ejection of most of the donor envelope largely independent of companion mass ratio. Line styles and additional quantities are defined as in Figure~\ref{fig:dem}.}
\label{fig:214q}
\end{figure*}

While the overall inspiral morphology remains qualitatively similar across the suite, increasing companion mass ratio leads to systematically stronger shocks, more rapid envelope expansion, and larger total ejecta masses as additional orbital energy and angular momentum are deposited into the donor during inspiral. However, the ratio of the simulated ejecta mass to the corresponding $\alpha=1$ energy-balance estimate does not increase systematically with $q$. This suggests that the additional orbital energy released by more massive companions is not converted solely into direct envelope unbinding, but is increasingly redistributed through shocks, large-scale hydrodynamic expansion, and internal restructuring of the donor, allowing the interaction to couple more efficiently to deeper and more tightly bound layers of the envelope.

The evolution of $\Psi$ remains qualitatively similar across the full range of mass ratios. In all cases, $\Psi$ increases substantially during the inspiral, indicating that the donor contracts as the embedded companion penetrates deeper into the envelope. These calculations therefore suggest that, for donors with modest central concentration, the companion mass ratio primarily regulates the efficiency of mass ejection while the overall character of the inspiral is determined by the donor structure itself. 

The high-central-concentration suite ($\rho_c/\bar{\rho}=962$) tells a markedly different story. In contrast to the $18M_\odot$ donor, all simulations involving the $2.14M_\odot$ donor exhibit a prolonged two-stage evolution across the full range of companion mass ratios (Figure~\ref{fig:214q}). The embedded companion initially undergoes rapid inspiral over several dynamical timescales, during which the donor contracts and $\Psi$ increases modestly above unity. This early plunge-in phase is followed by a transition to large-scale envelope expansion once the inspiral slows and the companion settles into a quasi-stationary orbit.

After this transition, the subsequent evolution becomes dominated by global hydrodynamic expansion rather than continued orbital decay. In all models, $\Psi$ decreases steadily as the donor envelope expands, driving sustained mass loss over tens to hundreds of dynamical timescales and ultimately ejecting most of the envelope largely independent of companion mass ratio.

While the overall hydrodynamic evolution remains qualitatively similar across the suite, increasing companion mass ratio primarily accelerates the envelope expansion and mass-loss evolution. More massive companions inject orbital energy and angular momentum into the donor more rapidly, generating stronger shocks and driving faster large-scale expansion, such that near-complete envelope ejection is reached on progressively shorter timescales.

In both donor suites, the tidal disruption radius increases with companion mass because more massive main-sequence companions have lower mean densities and are therefore disrupted at larger orbital separations within the donor envelope.

\begin{figure*}
\centering
\includegraphics[scale = 0.55]{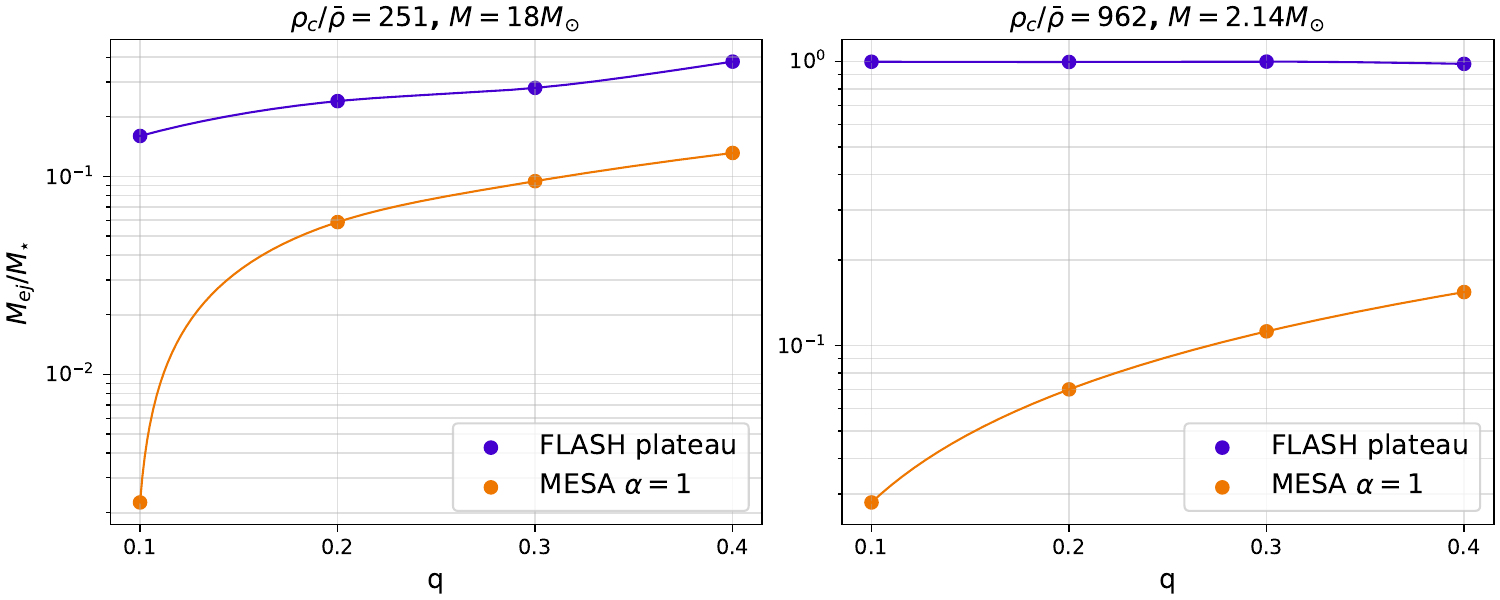}
\caption{Comparison between the total ejecta mass obtained from the hydrodynamical simulations and the predictions of the standard $\alpha$-formalism assuming $\alpha=1$. The {\it left} panel shows the moderately concentrated donor suite, while the {\it right} panel shows the high-central-concentration donor suite. The moderately concentrated donors broadly follow the orbital-energy scaling expected from the inspiral, although the simulations consistently eject more mass than predicted by the simple energy formalism. By contrast, the high-central-concentration donor suite ejects most of the donor envelope largely independent of companion mass ratio due to prolonged hydrodynamic expansion following the inspiral phase.}
\label{fig:mej}
\end{figure*}

\subsection{Hydrodynamic Mass Ejection Beyond the $\alpha$-Formalism}
\label{sec:massejection}

Figure~\ref{fig:mej} compares the total ejecta mass obtained from the hydrodynamical simulations to the predictions of the standard $\alpha$-formalism assuming $\alpha=1$. The left panel shows the moderately concentrated donor suite, while the right panel shows the high-central-concentration donor suite.

For the moderately concentrated donors, the total ejecta mass increases systematically with companion mass ratio and broadly follows the scaling expected from the orbital-energy budget. In these systems, the inspiral proceeds rapidly toward the donor core. As a result, orbital energy deposited after the inspiral passes the nominal $\alpha=1$ radius can continue to unbind progressively deeper layers of the donor envelope. However, the released energy is insufficient to eject all material exterior to the companion once the inspiral reaches the most tightly bound inner regions of the donor. The hydrodynamical simulations therefore consistently eject more mass than predicted by the simple energy formalism, while still retaining a clear dependence on the orbital energy released during inspiral. This behavior is broadly consistent with the findings of \citet{2024ApJ...977..196H}, who similarly found that hydrodynamic evolution can enhance envelope ejection beyond the predictions of simple instantaneous energy-balance arguments.

The comparison breaks down qualitatively for the high-central-concentration donor suite. In these systems, the simulations eject nearly the entire donor envelope over the full range of companion mass ratios, while the $\alpha=1$ formalism predicts substantially smaller ejecta masses. Once the inspiral settles into a quasi-stationary orbit, the subsequent evolution becomes dominated by large-scale hydrodynamic expansion of the donor envelope. In this regime, the donor itself performs much of the work required to unbind the envelope as expansion redistributes energy and drives prolonged mass loss over hundreds of dynamical timescales. The total ejecta mass is therefore no longer determined primarily by instantaneous orbital-energy deposition, but instead by the global hydrodynamic response of the donor. 

These conclusions may be altered substantially for main-sequence companions, whose lower mean densities make them susceptible to tidal disruption before reaching the smallest orbital separations within the donor envelope \citep{2026arXiv260210211T}. In such systems, the inspiral and associated orbital-energy deposition terminate once the companion is disrupted, potentially preventing the onset of the prolonged expansion-driven phase identified in the high-central-concentration donor suite. By contrast, compact companions such as white dwarfs, neutron stars, and black holes can survive to much smaller radii, allowing continued inspiral and sustained energy deposition deep within the donor envelope.

\section{Discussion, Lessons Learned, and Observational Implications} \label{sec:dis}

\subsection{Hydrodynamic Evolution in the Context of Common-Envelope Simulations}

The common-envelope literature has historically focused on highly extended RGB, AGB, and red-supergiant donors with radii spanning tens to hundreds of solar radii \citep{2012ApJ...746...74R,2012ApJ...744...52P,2016ApJ...816L...9O,2017MNRAS.464.4028I,2018MNRAS.477.3409I,apjac6269bib92}. As summarized in Appendix~\ref{cesim}, relatively few global simulations have explored the intermediate regime of mildly evolved donors with moderate central concentration considered here. Early hydrodynamical studies frequently employed idealized polytropic structures or simplified envelope models \citep[e.g.,][]{2012ApJ...744...52P}, whereas more recent calculations increasingly map one-dimensional stellar-evolution models directly into three-dimensional hydrodynamic simulations \citep{2012ApJ...746...74R,2016ApJ...816L...9O,2017MNRAS.464.4028I,2018MNRAS.477.3409I,apjac6269bib92}.

Despite substantial differences in numerical methods, resolution, and donor structure, these simulations reveal several common hydrodynamic features. In most calculations, the interaction begins with a rapid dynamical inspiral during which the embedded companion transfers orbital energy and angular momentum to the surrounding envelope through shocks, gravitational torques, and hydrodynamic drag. This initial plunge-in phase is typically followed by a substantial slowing of the inspiral as the local envelope around the companion becomes increasingly disturbed and partially evacuated. Many simulations subsequently develop long-lived oscillatory or quasi-periodic orbital evolution in which the companion continues orbiting within a partially bound envelope over many dynamical timescales \citep{2012ApJ...744...52P,2017MNRAS.464.4028I,apjac6269bib92}. These studies suggest that the outcome of the interaction is not determined solely by the orbital energy released during the initial plunge-in, but by the hydrodynamic state in which the inspiral ultimately stalls near the inner regions of the donor. Once the local envelope surrounding the companion becomes sufficiently disturbed, shocks, circulation flows, and large-scale envelope motions can redistribute deposited energy and angular momentum increasingly efficiently throughout the star \citep{2025ApJ...979L..11E}. The hydrodynamic response during this post-plunge evolution is often highly complex, with large-scale shocks, anisotropic outflows, envelope expansion, fallback, and recombination-driven cooling all contributing to the continued redistribution of energy and momentum throughout the envelope \citep{2013A&ARv..21...59I,2022MNRAS.512.4665L}. As a result, many simulations find that orbital-energy deposition during the initial dynamical inspiral alone is insufficient to unbind the full envelope, and that the subsequent hydrodynamic evolution over many dynamical timescales plays a critical role in determining the final ejecta mass and long-term fate of the system. However, these conclusions have largely emerged from simulations of highly extended giant envelopes, where the envelope dynamical timescales are long and the outer layers are only weakly bound.

\subsection{Hydrodynamic Evolution Regulated by Donor Central Concentration}
The simulations presented here explore this hydrodynamic evolution in a structurally distinct regime characterized by moderately evolved donors with central density contrasts of $\rho_c/\bar{\rho}\approx10^2-10^3$ (Figure~\ref{fig:1}). This range begins above the most centrally concentrated simple polytropic structures, but extends to density contrasts well below those of highly evolved giant envelopes. Within this intermediate regime, the donor central concentration emerges as a key parameter governing the global hydrodynamic response of the interaction \citep{2020ApJ...905..141L,2020ApJ...899...77E}. Additional comparison simulations presented in Appendix~\ref{comp} further support this interpretation by demonstrating that donors with substantially different masses and evolutionary states, but similar values of $\rho_c/\bar{\rho}$, exhibit qualitatively similar inspiral morphologies and mass-ejection histories . Importantly, these similarities emerge despite significant differences in the detailed stellar structure of the donor models, suggesting that $\rho_c/\bar{\rho}$ captures a physically important aspect of how the envelope couples hydrodynamically to the inspiraling companion. Systems with relatively modest central concentration, whose structures remain closer to polytropic envelopes, tend to evolve through a rapid inspiral toward the inner regions of the donor, with the evolution remaining largely dominated by local orbital-energy deposition during the dynamical plunge-in phase \citep[e.g.,][]{2012ApJ...744...52P}. By contrast, increasing central concentration leads to progressively stronger global envelope response as the inspiral slows and the local hydrodynamic drag around the companion weakens.

This transition toward slower, expansion-driven evolution bears some resemblance to the quasi-stationary outflow picture discussed by \citet{2011ApJ...731L..36I}, in which envelope ejection during common-envelope evolution may proceed through prolonged large-scale outflows rather than instantaneous dynamical unbinding. In our simulations, however, this behavior emerges naturally from the hydrodynamic coupling between the inspiral and the global envelope response, without requiring an explicit separation between instantaneous and stationary ejection phases. In this regime, shocks, circulation flows, and large-scale expansion can redistribute deposited energy and angular momentum more efficiently throughout the star (Figure~\ref{fig:hydro}).

The simulations explored here therefore appear to probe a progression in hydrodynamic behavior with increasing donor central concentration. At low concentration, the evolution is dominated by rapid inspiral driven by strong local drag with comparatively limited global envelope response (Figure~\ref{fig:18q}). At intermediate concentration, the slowing inspiral allows increasingly efficient hydrodynamic communication throughout the donor, driving large-scale expansion and efficient envelope ejection (Figure~\ref{fig:214q}). At still larger central concentration, the development of a strongly condensed inner structure increasingly decouples the inspiral from the outer envelope, reducing communication between the inner and outer regions of the star and returning the interaction toward the more classical extended-envelope regime in which the final outcome depends sensitively on where the companion ultimately parks within the donor \citep[e.g.,][]{2016ApJ...816L...9O}.

A final important caveat is that essentially all of the hydrodynamic regimes discussed here assume that the embedded companion behaves as a point mass throughout the interaction. In reality, the survival of the companion may itself depend sensitively on its internal structure and compactness \citep[e.g.,][]{2026arXiv260210211T}. Extended main-sequence companions may undergo tidal disruption or substantial mass loss before reaching the deepest stages of inspiral, potentially altering or even terminating the subsequent hydrodynamic evolution (Figures~\ref{fig:18q} and \ref{fig:214q}). Consequently, the mass-ejection histories and large-scale envelope responses identified here may not only reflect the structure of the donor, but, in systems where the interaction appears to terminate prematurely, could also provide indirect constraints on the compactness and survivability of the embedded companion.

\subsection{Implications for Envelope Ejection and Luminous Red Novae}
The current understanding of luminous red novae increasingly favors a picture in which the observed diversity of LRNe reflects the complex and time-dependent hydrodynamic evolution of the binary interaction itself rather than a single universal ejection mechanism \citep{2026arXiv260517005K}. In particular, observations increasingly point toward prolonged pre-maximum mass loss, multiple ejection episodes, and extended interaction between successive outflows, implying that the envelope response can remain dynamically active over many orbital or dynamical timescales. This emerging observational framework aligns naturally with the hydrodynamic complexities we have identified for donors spanning different evolutionary states and central density concentrations. 

The simulations presented here suggest that the time dependence of mass ejection may be as important observationally as the total ejecta mass itself. In systems with relatively low central density concentration, where the evolution is dominated by rapid inspiral and strong local drag \citep{2015ApJ...803...41M,2017ApJ...838...56M,2023ApJ...954..176Y,2024ApJ...977...16R}, the envelope response remains closely tied to the instantaneous release of orbital energy, producing comparatively impulsive mass ejection on dynamical timescales, similar to the rapid plunge-in behavior seen in lower-concentration or polytropic-envelope simulations \citep{2012ApJ...744...52P,2012ApJ...746...74R}. At intermediate central concentration, however, the slowing inspiral allows increasingly efficient hydrodynamic communication throughout the donor, leading to prolonged expansion-driven evolution in which the envelope itself continues driving mass loss long after the rapid plunge-in phase has slowed substantially (Figure~\ref{fig:214q}). In this regime, the outflow transitions from a predominantly impulsive ejection event toward a more sustained wind-like or multi-stage mass-loss history. 

The importance of the mass-ejection history is also emphasized by LRN light-curve modeling, which shows that the observable properties of these transients depend sensitively on the ejecta mass, velocity, launching radius, and the interaction between dynamically ejected material and pre-existing outflows \citep{2022ApJ...938....5M}. At still larger central concentration, corresponding to the highly extended giant envelopes explored in many previous common-envelope simulations \citep{2016ApJ...816L...9O,2017MNRAS.464.4028I,2018MNRAS.477.3409I,apjac6269bib92,2022MNRAS.512.4665L}, the inspiral timescale again becomes long compared to the local dynamical time of the envelope, and the evolution returns to a regime in which envelope ejection proceeds gradually over many dynamical timescales before the companion reaches the deepest layers of the donor \citep{2022ApJ...937...96M,2026arXiv260210211T}.

This progression in hydrodynamic behavior may therefore provide a natural framework for understanding the diversity of luminous red nova outbursts \citep{2022ApJ...938....5M,2026arXiv260517005K,2026arXiv260210211T}. In this picture, the observable properties of LRNe depend not only on the total ejecta mass \citep[e.g.,][]{2017ApJ...835..282M}, but also on the temporal structure of the outflow itself, including whether the interaction produces an impulsive shell, a prolonged wind-like phase, or multiple successive ejection episodes. Systems in the intermediate regime identified here may be particularly well suited for generating long-lived outbursts, extended photospheric evolution, and complex circumstellar environments powered by continued shocks and interaction between successive phases of mass ejection.

\appendix

\begin{figure*}
\centering
\includegraphics[scale = 0.67]{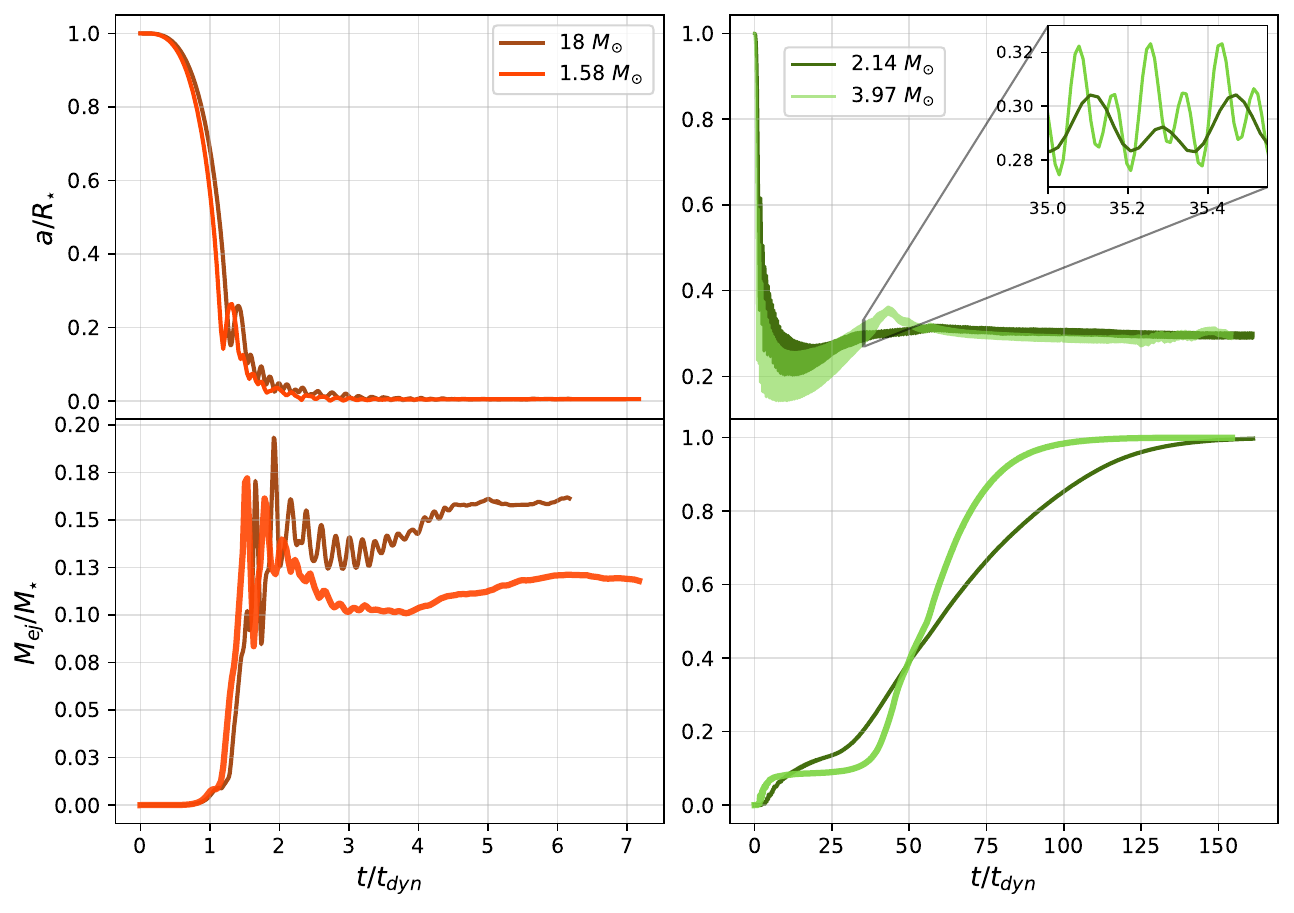}
\caption{Comparison simulations involving donors with similar central density contrasts but substantially different masses and evolutionary states. Top panels show the orbital separation evolution normalized by the donor radius, while bottom panels show the cumulative unbound mass fraction normalized by the total donor mass as a function of time normalized by the donor dynamical timescale. Despite significant differences in stellar mass and detailed envelope structure, systems with similar values of $\rho_c/\bar{\rho}$ exhibit qualitatively similar inspiral morphologies and mass-ejection histories, supporting the interpretation that donor central concentration regulates the global hydrodynamic response during common-envelope evolution.}
\label{fig:scs}
\end{figure*}

\section{Comparison Simulations at Fixed Stellar Central Concentration}\label{comp}

To further isolate the role of donor central concentration, we compare simulations involving stars with substantially different masses and evolutionary states but similar values of $\rho_c/\bar{\rho}$. Figure~\ref{fig:scs} compares the inspiral evolution and cumulative mass ejection for two such pairs of systems: a $1.58M_\odot$ donor with $\rho_c/\bar{\rho}\approx 260$ compared to the $18M_\odot$ donor, and a $3.97M_\odot$ donor with $\rho_c/\bar{\rho}\approx940$ compared to the $2.14M_\odot$ donor. All models shown correspond to the $q=0.1$ simulations.

Despite large differences in donor mass, radius, and detailed stellar structure, systems with similar values of $\rho_c/\bar{\rho}$ exhibit remarkably similar hydrodynamic evolution. In the lower-central-concentration comparison pair ($18M_\odot$ and $1.58M_\odot$), both systems undergo continued inspiral toward the inner regions of the donor with comparatively limited prolonged expansion-driven evolution. By contrast, the higher-central-concentration pair ($2.14M_\odot$ and $3.97M_\odot$) both transition into quasi-stationary orbital evolution accompanied by sustained expansion-driven mass loss after the rapid plunge-in phase. In both comparison pairs, the overall inspiral morphology and character of the mass-ejection history closely resemble one another once normalized by the donor dynamical timescale, stellar radius, and total mass.

Importantly, realistic stellar models with similar values of $\rho_c/\bar{\rho}$ are not structurally self-similar. Unlike polytropic stellar models, which are self-similar by construction, realistic stellar evolution models with comparable values of $\rho_c/\bar{\rho}$ can possess substantially different density gradients, entropy profiles, and envelope structures. The similarity of the resulting hydrodynamic evolution therefore suggests that the central-to-average density ratio captures a physically important aspect of how the donor envelope couples to the inspiraling companion and redistributes deposited energy and angular momentum during the interaction.

The most significant differences between the comparison models appear primarily in the late-time mass-ejection histories. In particular, the more centrally concentrated systems develop prolonged expansion-driven evolution that can continue ejecting mass long after the rapid inspiral phase has slowed substantially. Nevertheless, the overall qualitative agreement between systems with similar $\rho_c/\bar{\rho}$ supports the interpretation that donor central concentration acts as a key organizing parameter governing the global hydrodynamic response during common-envelope evolution.

\begin{figure*}
\centering
\includegraphics[scale = 0.5]{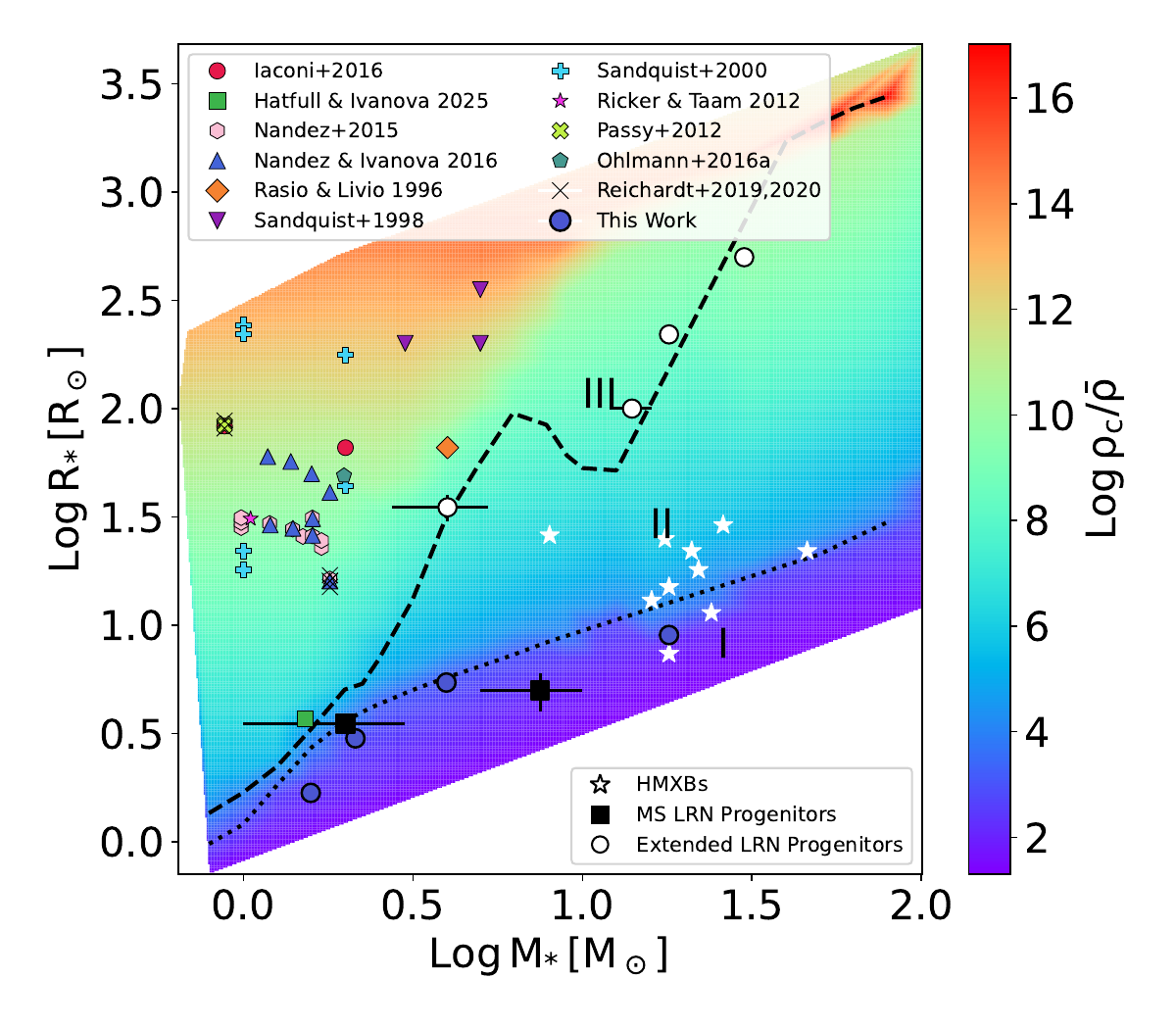}
\caption{Initial donor masses and radii adopted in previous global common-envelope simulations compared to the donor models studied in this work ({\it blue circle} symbols). Literature simulations primarily probe evolved RGB, AGB, and red-supergiant donors with large radii, while the present models occupy a distinct regime of mildly evolved progenitors with intermediate central concentration. The masses and radii of eclipsing high-mass X-ray binaries studied by \citet{2015A&A...577A.130F} are shown as white star symbols. The background color indicates the central-to-average density ratio, $\rho_c/\bar{\rho}$, highlighting that the present study explores a region of donor structure that has been comparatively underrepresented in previous three-dimensional common-envelope calculations. The dotted and dashed lines delineate the three evolutionary regimes identified by \citet{2026arXiv260210211T}. Binary population demographics based on the \citet{Sana2012} initial binary distributions suggest that approximately 27.4\%, 11.7\%, and 58.2\% of systems undergoing unstable mass transfer populate Zones I, II, and III, respectively. Observed LRN progenitors with pre-explosion imaging are plotted for comparison \citep{Matsumoto2022}. From left to right, in order of increasing inferred mass, these include V1309 Scorpii, M31LRN2015, V838 Monocerotis, M101-OT2015, SN Hunt248, and AT~2018bwo. These progenitors are either main-sequence progenitors ({\it black squares}; Zone I) or extended progenitors ({\it white dots}; Zones II and III).}
\label{fig:ce_literature_mr}
\end{figure*}
\section{Donor Structures in Previous Common-Envelope Simulations}\label{cesim}
The initial donor masses and radii adopted in previous global common-envelope simulations are summarized in Figure~\ref{fig:ce_literature_mr}. Existing calculations have predominantly focused on evolved RGB, AGB, and red-supergiant donors with large radii and strongly stratified envelopes \citep{1996ApJ...471..366R,2000NewA....4..313S,2012ApJ...744...52P,2012ApJ...746...74R,2015ApJ...806..170N,2016MNRAS.460.3992N,2017MNRAS.464.4028I,2018MNRAS.477.3409I,apjac6269bib92,apjac6269bib91}. By contrast, the mildly evolved progenitors studied in this work occupy a comparatively underexplored region of parameter space, with smaller radii at fixed mass and intermediate central concentration.

This distinction is important because the hydrodynamic response to inspiral depends not only on the total envelope binding energy, but also on how the donor structure regulates the redistribution of deposited energy and angular momentum during and after the dynamical plunge-in phase. In highly extended giant envelopes, the inspiral often evolves toward a regime in which the companion becomes increasingly decoupled from the outer envelope as the local density rises and the inspiral slows near the condensed inner regions of the donor. By contrast, the moderately evolved progenitors considered here remain sufficiently coupled to the global envelope. 

The background color in Figure~\ref{fig:ce_literature_mr} indicates the central-to-average density ratio, $\rho_c/\bar{\rho}$, illustrating how the present models bridge the transition between low-concentration structures and the highly stratified envelopes of classical giant-star common-envelope simulations. The intermediate regime studied  here appears particularly favorable for prolonged expansion-driven mass loss and efficient global hydrodynamic response, phenomena that are less apparent in both the low-concentration rapid-inspiral limit and the highly stratified giant-envelope regime. The figure also highlights the three qualitative evolutionary  regimes identified by \citet{2026arXiv260210211T}.

\begin{acknowledgments}
 We gratefully acknowledge helpful discussions with A. Antoni, R. Everson, A. Vigna-G\'omez, A. Rosselli-Calderon, R. Yarza, M. MacLeod, A. Villar, and J. Law-Smith. This research was supported by the Lamat Institute \citep{2025NatAs...9.1770Q}, the Heising-Simons Foundation, and NSF grants AST-1852393, AST-1911206, AST-2150255, and AST-2206243 through UC Santa Cruz. We acknowledge use of the \textit{lux} supercomputer at UCSC, funded by NSF MRI grant AST-1828315, as well as the HPC facility at the University of Copenhagen, funded by VILLUM FONDEN (project number 16599). The authors used OpenAI Codex \citep{OpenAI2025_Codex} for assistance with manuscript editing.
\end{acknowledgments}

% \software{FLASH \citep{Fryxell2000},
%     Python,
% 	MESA \citep{Paxton2011, Paxton2013, Paxton2015, Paxton2018, Paxton2019}, 
% 	matplotlib \citep{Hunter2007},
%     yt \citep{2011ApJS..192....9T},
% 	NumPy \citep{vanderwalt2011}, 
% 	py\_mesa\_reader \citep{WolfSchwab2017}}

\bibliography{bibliography}
\end{document}